\documentclass[twocolumn,preprintnumbers,superscriptaddress,amsmath,amssymb,aps,showpacs]{revtex4}
\usepackage{graphicx,float}

\begin{document}

\title{Non-Poisson Renewal Events and  Memory}


  \author{Rohisha Tuladhar}
 \email{raisha.t@gmail.com}
 \affiliation{Center for Nonlinear Science, University of North Texas,
 	P.O. Box 311427, Denton, Texas 76203-1427}
 
\author{Mauro Bologna}
 \affiliation{Instituto de Alta Investigaci\'{o}n, Universidad de Tarapac\'{a}, Casilla 6-D, Arica, Chile}
 
 \author{Paolo Grigolini}
 \email{Paolo.Grigolini@unt.edu}
 \affiliation{Center for Nonlinear Science, University of North Texas,
 	P.O. Box 311427, Denton, Texas 76203-1427}

\begin{abstract}
We study two different forms of  fluctuation-dissipation  processes generating 
 anomalous relaxations to equilibrium of an initial out of equilibrium condition, the former being based on a stationary although very slow correlation function and the latter 
characterized by the occurrence of crucial events, namely,  non-Poisson renewal events, incompatible with the stationary condition. Both forms of regression to equilibrium have the same non-exponential Mittag-Leffler structure.  We analyze the  single trajectories of the two processes by recording the time distances between two consecutive origin re-crossings and establishing the corresponding waiting time probability  density function (PDF), $\psi(t)$. In the former case, with no crucial events, $\psi(t)$ is exponential and in the latter case, with crucial events, $\psi(t)$ is an inverse power law PDF with a diverging first moment.  We discuss the 
consequences that this result is expected to have for the correct interpretation of some anomalous relaxation processes.  

\end{abstract}

\pacs{05.40.-a, 05.40.Fb, 05.10.Gg, 05.30.Pr} 
\maketitle

\section{Introduction} \label{intro}

Exponential relaxation is a popular signature of conventional statistical physics.  In the last years a form of non-exponential relaxation attracting the attention of the researchers in the field of complexity
has been the Mittag-Leffler (ML) relaxation \cite{ml}.  Metzler and Klafter \cite{metzler} made the interesting observation that the ML relaxation function $E_{\alpha}(-(\lambda t)^{\alpha})$, with $\alpha < 1$,  has the remarkable 
property of being proportional to the stretched exponential function $exp(-(\lambda t)^{\alpha})$ in the time region $t < 1/\lambda$ and to the inverse power law (IPL) $1/t^{\alpha}$ in the time region $t > 1/\lambda$. If $\lambda \ll 1$ the initial time region may be very extended and this property,  according to Metzler and Klafter,  establishes a bridge between two conflicting parties in the field of dielectric relaxation, namely between the advocates of stretched exponential functions and the advocates of IPL's.  The interest for ML relaxation is growing and it extends to several fields of investigation, from diffusion in 
biological tissue \cite{cardiac} to dielectric relaxation \cite{coffey,metzler,mainardi} and from chemical reactions \cite{chemicalreactions} to neural dynamics \cite{neurons}. It is also important to stress the importance of the ML exponential function for 
the definition of fractional derivative in time \cite{mainardi2} and for the related problem of  interpreting the Continuous Time Random Walk (CTRW) \cite{montrollweiss}  as the representation in the clock time of the ordinary diffusion occurring in the operational time \cite{pramukkul}.

What is the physical origin of the ML relaxation? Is the ML relaxation compatible with a Hamiltonian picture?  It is well known (see, for  example \cite{vitali}) that a rigorous Hamiltonian approach to relaxation 
yields significant deviations from the exponential relaxation. The Generalized Langevin Equation (GLE) \cite{mori,lastzwanzig} is known to generate exponential relaxation under strong approximations, called Markov approximations. It is not quite surprising that the GLE may generate the ML non-exponential form of relaxation.  In  fact, in  2011 Pottier \cite{pottier} proved that the GLE can be assigned a suitable memory kernel yielding for the regression to equilibrium of the variable driven by the GLE the ML non-exponential behavior.  More recently, Kneller \cite{kneller2,kneller} adopted the same \emph{first principle}  approach  as that used by Pottier to study the autocorrelation function for a  solute particle slowly diffusing in a bath of fast solvent molecules that generate, however, cooperation and consequently slow fluctuations, preventing the Markov approximation from turning the GLE into an ordinary Langevin equation. The Mori-Zwanzig GLE, as pointed out by Kneller,  is conceptually different from the stochastic GLE of Ref. \cite{stochasticGLE}. In fact, the Mori-Zwanzig approach is derived from a fully Hamiltonian picture,  while the authors of
Ref. \cite{stochasticGLE}, although using the same generalized fluctuation-dissipation structure as the Mori-Zwanzig GLE, adopt for the fluctuation the Fractional Gaussian Noise (FGN) that generates the Fractional Brownian Motion (FBM) diffusion \cite{fbm}, when dissipation is neglected.  The key property of both forms of GLE is that the time derivative of the variable of interest that we call $x(t)$,  is a time convolution structure between the memory kernel and $x(t-t')$, with $t' <t$. This is the reason why the variable $x(t)$ is thought to have memory: its time evolution from time $t$ onwards depends on the past history of $x(t)$.  Note that in this paper we denote the variable of interest with the symbol $x(t)$ rather than $v(t)$, which would be appropriate for the case when the variable of interest is indeed a velocity. We adopt the symbol $x(t)$ to stress the generality of our approach and to facilitate the applications of  the results of this paper to a wider set of processes.

In this paper we address the issue of comparing the GLE approach to ML relaxation  to another frequently adopted theoretical approach to ML relaxation: the subordination approach. The subordination approach to ML relaxation is based on the assumption that in the so called operational time the $x$ trajectories are driven by the conventional Langevin equation and that the time evolution of corresponding probability density function (PDF)  $p(x,t)$ is determined by the ordinary Fokker-Plank equation. In this paper we refer to the operational events, perceived in the 
clock-time scale, as \emph{crucial events} responsible for the system time evolution. In the clock-time representation there are no events in the extended time intervals between two consecutive crucial events. In this case the 
non-Markovian structure of the GLE, namely the time convolution between memory kernel and the time evolution of the variable of interest, is replaced by the time convolution between a memory kernel and 
the  function $\mathcal{L}_{FP} p(x,t-t')$, where $\mathcal{L}_{FP}$ denotes the ordinary Fokker-Plank operator defined in this paper by Eq. (\ref{definition1}). This property suggests that the subordination approach may lead to the same memory properties as the GLE theoretical approach. 

It is important to stress that this subordination approach to ML relaxation is shared by many authors, even if this connection is not immediately evident. The work of Ref. \cite{metzlerfractional} with the structure 
\begin{equation} \label{introduction}
\frac{\partial}{\partial t} p(x,t)  =  \frac{\partial^{1-\alpha}}{\partial t^{1-\alpha}} \mathcal{L}_{FP} p(x,t),
\end{equation}
is virtually equivalent to
\begin{equation} \label{introduction2}
\frac{\partial^{\alpha}}{\partial t^{\alpha}} p(x,t)  =   \mathcal{L}_{FP} p(x,t),
\end{equation}
as it can be easily understood by applying to both sides of Eq. (\ref{introduction}) the fractional derivative 
$\frac{\partial^{\alpha -1}}{\partial t^{\alpha -1}}$. This is not a rigorous demonstration because the fractional derivative on the right hand side of Eq. (\ref{introduction}) is the Riemann-Liouville fractional derivative \cite{physicsREPORT} and the fractional derivative on the left hand side of Eq. (\ref{introduction2})  is the Caputo fractional derivative \cite{pramukkul}.  However, this simple heuristic argument leads to the correct physical interpretation of both Eq. (\ref{introduction}) and Eq. (\ref{introduction2}) and to the important conclusion that the emergence of ML relaxation out of them is based on the subordination perspective, as discussed in detail in Ref. \cite{pramukkul}. For an earlier discussion the readers can consult also the work of 
\cite{earlierSOKOLOV}.  In this sense also Ref. \cite{metzlerfractional}, as well Ref. \cite{physicsREPORT}, is based on the subordination perspective. It is worth remarking that the adoption of the 
fractional derivative structure of Eq. (\ref{introduction}) was used to generalize the Kramers equation \cite{mixture} thereby leading to a mixture picture such as that of \cite{coffey} that cannot be directly connected to subordination. 

The main purpose of this paper is to focus on the derivation of 
ML relaxation on the basis of either the GLE theoretical perspective or the subordination approach. In the former case 
the relaxation is based on a slow but stationary correlation function and in the latter case it depends on the occurrence of 
crucial events. Both approaches lead to the same relaxation to equilibrium of $<x(0)> \neq 0$, but the single trajectories of the former case
are characterized by a behavior quite different from that of the trajectories of the latter case. To establish the striking difference between the single trajectories of the former case and the single trajectories of the latter case, we record the times of origin crossing in both cases and the time intervals between two consecutive crossings, called \emph{permanence times}.  We evaluate the waiting time PDF $\psi(t)$, with $t$ being the time interval between two consecutive crossings.  According to ordinary statistical physics one would expect
\begin{equation} \label{apparentlyordinary} 
\psi(t) = r exp(-rt),
\end{equation}
in striking conflict with the signature of complexity given by
\begin{equation} \label{complexity}
\psi(t) \propto \frac{1}{t^{\mu_R}}.
\end{equation}
In this paper we prove that GLE generates  the IPL behavior of Eq. (\ref{complexity}) in the short-time regime and the exponential behavior of Eq. (\ref{apparentlyordinary}) in the long-time limit, if the stretched exponential regime of the ML relaxation is very extended. When the stretched exponential regime of ML relaxation is negligible the GLE  yields only the exponential regime of
Eq. (\ref{apparentlyordinary}). 
The subordination approach to ML relaxation generates a completely different behavior for $\psi(t)$.  It generates the IPL regime of Eq. (\ref{complexity}) regardless of whether the stretched exponential regime of the ML relaxation is very extended or completely negligible. However, when the ML relaxation is characterized by an extended stretched exponential regime the index $\mu_R$ is significantly smaller than the complexity index $\mu_R$ generated by a ML relaxation lacking the stretched exponential regime.  Notice that working with time series of a finite length generates an exponential truncation of $\psi(t)$ also in the subordination case. However, the length of the IPL regime in this case can be increased by increasing the length of the observed time series, while in the GLE case the exponential truncation does not depend on the length of the observed time series.  This is a real physical property, proved by an exact analytical theory, generated by extended memory that establishes a correlation between the permanence times $t$.

The outline of the paper is as follows. In Section \ref{review1}
we review the two distinct ways of generalizing the ordinary 
process of fluctuation-dissipation discussed in this paper. 
In Section \ref{GLE} we review the approach to ML relaxation based on the GLE theoretical perspective. Section \ref{thesameinitialcondition} shows why the regression to equilibrium of the non-vanishing initial condition $<x(0)>$ based on subordination is identical to that given by GLE theoretical approach, and consequently yields the ML relaxation. Section \ref{trajectories} illustrates the original results of this paper, namely, that the fluctuations of the single trajectories around the origin in the GLE case are described by an exponential waiting time PDF,  whereas 
in the subordination case are described by an IPL waiting time PDF. Finally we devote Section \ref{last} to concluding remarks and to a plan for the applications of the results of this paper.

\section{A traditional way to go beyond ordinary fluctuation-dissipation processes} \label{review1}

The GLE for a stochastic variable $x$ is given by the following time convoluted structure \cite{grigoacp}:

\begin{equation} \label{gle}
\frac{d}{dt} x = - \int_{0}^t dt' \varphi(t') x(t-t') + \xi(t) .
\end{equation}
The memory kernel $\varphi(t)$ is related to the stationary and normalized correlation function of $\xi(t)$, 
\begin{equation}
\Phi_{\xi}(\tau) = \frac{\left<\xi(t+\tau) \xi(t) \right>}{\left<\xi(t)^2\right>}  ,
\end{equation}
with the independence of absolute time $t$ stressed by mean of the notations
\begin{equation}
\left<\xi(t+\tau) \xi(t) \right> = \left<\xi(\tau) \xi \right>_{eq} 
\end{equation}
and
\begin{equation}
\left<\xi(t) \xi(t) \right> = \left<\xi^2 \right>_{eq} .
\end{equation}
The memory kernel  $\varphi(t)$ is related to $\Phi_{\xi}(\tau)$ by 
\begin{equation} \label{fromphitoPhi}
\varphi(t) = \Delta^2 \Phi_{\xi}(t),
\end{equation}
with $\Delta^2$ denoting the intensity of the coupling between the variable $x$ and the variable $\xi$. 
The notation $\left< ...\right>$ is used throughout this paper to denote ensemble averages. 

The variables $x(t)$ and $\xi$ are assumed to obey the equilibrium condition
\begin{equation}
\left<x\right>_{eq} = 0
\end{equation}
and
\begin{equation} \label{bath}
\left<\xi\right>_{eq} = 0.
\end{equation}

When an initial out of equilibrium condition $\left<x(0) \right>  \neq 0$ is realized, the regression to equilibrium is obtained from Eq. (\ref{gle}) by making an ensemble average that thanks to Eq. (\ref{bath}) yields
\begin{equation} \label{ensemble1}
\frac{d}{dt} \left<x(t)\right>  = - \int_{0}^t dt' \varphi(t') \left< x(t-t')\right> .
\end{equation}

Let us now consider the following time-convoluted Fokker-Planck equation
\begin{equation} \label{timeconvoluted}
\frac{\partial}{\partial t} p(x,t)   = \int_{0}^t dt' \varphi(t') \mathcal{L}_{FP} p(x,t-t'),
\end{equation}
where  $\mathcal{L}_{FP}$ is the  dimensionless Fokker-Planck operator
\begin{equation} \label{definition1}
\mathcal{L}_{FP}  \equiv   \left\{\frac{\partial}{\partial x}x + \left<x^2\right>_{eq} \frac{\partial^{2}}{\partial x^{2}}\right\} 
 \end{equation}

Using the method of integration by parts it is straightforward to prove that this generalized Fokker-Planck equation yields the 
same regression to equilibrium as the GLE of Eq. (\ref{gle}), namely, Eq. (\ref{ensemble1}). This is a reasonable property if we take into account that
the condition 
\begin{equation}
\varphi(t) =  2\omega \delta(t)
\end{equation} 
turns Eq. (\ref{gle}) into
\begin{equation} \label{le}
\frac{d}{dt} x = - \omega x(t) + \xi(t)
\end{equation}
and Eq.(\ref{timeconvoluted})
into
\begin{equation} \label{fp}
\frac{\partial}{\partial t} p(x,t)  = \omega \left\{\frac{\partial}{\partial x}x + \left<x^2\right>_{eq} \frac{\partial^{2}}{\partial x^{2}}\right\} p(x,t),
\end{equation} 
namely the standard Langevin equation and its equivalent probabilistic representation, the standard Fokker-Planck equation. 
In this article, we focus our attention on the case where $\varphi(t)$ has a negative long-time tail.

Eq. (\ref{timeconvoluted}) may be interpreted as the PDF representation corresponding to the GLE of Eq. (\ref{gle}). However, it  is not so. The correct Fokker-Planck equation corresponding to the GLE of Eq. (\ref{gle}) was found in 1976 by Adelman \cite{adelman}. In the case when the friction is neglected and only diffusion is taken into account the correct PDF representation of the process of Eq. (\ref{gle}) is given by \cite{grigolini}
\begin {equation}
\frac{\partial}{\partial t} p(x,t) = Q(t) \frac{\partial^2}{\partial x^2}p(x,t),
\end{equation}
where
\begin{equation} \label{rohishanotation}
Q(t) \equiv \frac{\left<\xi^2\right>_{eq}}{\Delta^2} \int_{0}^t dt' \varphi(t').
\end{equation}
As we see in Section \ref{thesameinitialcondition}, Eq. (\ref{timeconvoluted}) has a physical origin totally different from that of Eq. (\ref{gle}) characterized by the occurrence of crucial events that are not present in Eq. (\ref{gle}).

\section{Anomalous Relaxation using GLE} \label{GLE}
The GLE picture that we adopt in this article rests on the non-Ohmic bath picture \cite{weiss}, where

\begin{equation} \label{wanted} 
\Phi_{\xi}(t)  \approx \textrm{sign}(1-\delta) \frac{a}{t^{\delta}}, \,\,\, t\rightarrow \infty,
\end{equation}
with $0 < \delta < 2$ and $a$ a kind of normalization constant. The scaling of the diffusion process generated by the fluctuation $\xi(t)$ when $\Delta^2 = 0$ is denoted with the symbol $H$, called Hurst coefficient, and is related to $\delta$ by the relation \cite{cakirarkadi}
\begin{equation} \label{cakirrelation}
H = 1 - \frac{\delta}{2}.
\end{equation}
Thus the condition $1 < \delta < 2$ corresponds to $H < 0.5$, sub-diffusion, and the condition $0< \delta < 1$ corresponds to $H > 0.5$, super-diffusion.  We show that the first condition yields a ML function with 
$\alpha < 1$, which is the main focus of this paper. However, for further information we explore also the case $0< \delta < 1$, yielding, as shown hereby, the  ML function with $\alpha > 1$. 
In  this paper we adopt for  Laplace transform the notation:
\begin{equation}
\hat f(u) = \mathcal{L} \left\{f(t)\right\} = \int_{0}^{\infty} dt
\exp(-ut) f(t) dt.
\end{equation}
It is convenient to remind the readers that the Laplace transform of the ML function $E(t)$ is
\begin{equation} \label{laplacetransformofml}
\hat E(u) = \frac{1}{u + \lambda^{\alpha} u^{1-\alpha}},
\end{equation}
with $0 < \alpha < 2$.

\subsection{Moving from subdiffusion}

For a super-Ohmic bath, $1< \delta < 2$, the anti-correlation negative tail must be ``compensated" by the positive values of  $\Phi_{\xi}(t)$ at short times in order to provide the necessary condition for sub-diffusion \cite{rasit}
\begin{equation}
\label{sub}
\int_0^{\infty} \Phi_{\xi}(t) dt = 0.
\end{equation}
In this paper we focus on the
long-time limit, and both  the numerical calculations and the theoretical discussion are done with a
correlation function $\Phi_{\xi}(t)$ fulfilling the condition (\ref{sub}) and also $\Phi_{\xi}(0)=1$. The analytical expression is

 \begin{equation} \label{weneedmauro}
 \Phi_{\xi}(t) = \frac{1}{2-\delta} e^{-\gamma t} - \frac{\delta-1}{2-\delta} \frac{1}{\left(1 + \gamma t \right)^{\delta}}, \end{equation}
yielding for the normalization constant the 
value
\begin{equation} \label{adefined}
 a = \frac{\delta -1}{(2-\delta) \gamma^{\delta}}.
\end{equation}
The numerical work, as done on the earlier paper of Ref. \cite{elvis}, is realized generating first the free FBM diffusion $x(t)$ using the algorithm of  \cite{algorithm}  and deriving 
the FGN $\xi(t)$ from it by time differentiation, $\xi(t) \equiv dx/dt$. This is equivalent to assigning to the parameter $\gamma$ in Eq. (\ref{weneedmauro} the maximum possible value, which is of the order of unity, taking into account that the integration time step is $\Delta t = 1$.

The Laplace transform of  $\Phi_{\xi}(t)$ of Eq. (\ref{weneedmauro})  is

\begin{equation} \label{use}
\hat \Phi_{\xi}(u) =  \frac{1}{2-\delta} \frac{1}{\gamma + u} - \frac{\delta -1}{2-\delta} \frac{1}{\gamma^{\delta}} \Gamma(1-\delta) u^{\delta-1}.
\end{equation}
The time convolution structure of  Eq.(\ref{ensemble1}) makes it easy to express the Laplace transform of $\left<\hat x(u)\right>$ in terms of the Laplace transform of the memory kernel
$\varphi(t)$, which, due to Eq. (\ref{fromphitoPhi}), requires the use of Eq. (\ref{use}).  This leads to
\begin{equation} \label{rohishasuggestion}
\left<\hat x(u)\right> = \frac{1}{u+ \Delta^2 \Phi_{\xi}(u)} \left< x(0)\right> .
\end{equation}
Note that the fast transition of the correlation function $\Phi_{\xi}(t)$ of Eq. (\ref{weneedmauro}) does not affect the long-time behavior of the system \cite{pottier}.  According to Pottier \cite{pottier2} the first term on the right hand side of Eq. (\ref{use}) is neglected because it generates a singularity to the left of the integration path to do when inverting the Laplace transform through integration on the Bromwich contour.

In conclusion, by making the Laplace transform of Eq.~(\ref{ensemble1}) we obtain
\begin{equation} \label{antido}
\left<\hat x(u)\right>  = \frac{1}{u +\lambda^{2-\delta}
u^{\delta - 1}} \left<x(0)\right>,
\end{equation}
where
\begin{equation} \label{lambdadefined}
\lambda^{2-\delta} = \frac{\pi \Delta^2 a}{\Gamma(\delta)\sin\left(\pi (\delta-1)\right)}.
\end{equation}
Inverse Laplace transforming Eq.~(\ref{antido}) we get the solution for the average coordinate
\begin{equation} \label{ml}
\frac{\left<x(t)\right>}{\left<x(0)\right>}=E_{2-\delta}(-(\lambda
t)^{2-\delta}),
\end{equation}
where
\begin{equation}
\label{mg}
E_{\alpha}(z) = \sum_{k=0}^{\infty} \frac{z^k}{\Gamma(\alpha k +1)}
\end{equation}
is the ML function, as found by Pottier \cite{pottier2}. It follows from the properties of the ML function that the relaxation of $\left<x(t)\right>$ occurs exponentially only for the $\delta = 1$, otherwise it is a slower process in the case of super-Ohmic bath, $1< \delta < 2$.

Note that the variable $x(t)$ driven by Eq. (\ref{gle}), with the fractional Gaussian noise  $\xi(t)$ corresponding to the FBM power index $H < 0.5$ is actually another fractional Gaussian noise with
\begin{equation} \label{veryimportant}
H' = 1 - H.
\end{equation}
To prove this important property, we use the Onsager principle, 
namely the assumption that the decay of the equilibrium correlation function is identical to the regression to equilibrium of an out of equilibrium condition,
\begin{equation} \label{onsager}
\Phi_{x}(t) = \frac{\left<x(t)\right>}{\left<x(0)\right>}.
\end{equation}
We refer the reader to Ref. \cite{onsager} for an earlier use of this method.
Using Eq. (\ref{onsager}) we rewrite Eq. (\ref{rohishasuggestion}) as
\begin{equation} \label{rohishasuggestion2}
\hat \Phi_x(u) = \frac{1}{u+ \Delta^2 \hat \Phi_{\xi}(u)} .
\end{equation}
Using Eq. (\ref{antido}) we write Eq. (\ref{rohishasuggestion2}) as
\begin{equation} \label{antido2}
\hat \Phi_{x}(u)  = \frac{1}{u +\lambda^{2-\delta}
u^{\delta - 1}}.
\end{equation}
This equation is made identical to the ML function of Eq. (\ref{laplacetransformofml})  by setting 
\begin{equation}
\alpha = 2 - \delta.
\end{equation}
Note that 
\begin{equation}
1 < \delta < 2 ,
\end{equation}
thereby yielding $\alpha < 1$. Note also that in the numerical calculations of Section \ref{numerical1} the minimal value of the elementary time step is $\Delta t = 1$, thus implying that $u < 1$. This yields 
$u^{\delta-1} > u$.  As a consequence when $\Delta^2$ is large enough as to make $\lambda$, as determined by Eq. (\ref{lambdadefined}),  of the order of unity, we can rewrite Eq. (\ref{antido2}) as
\begin{equation} \label{antido3}
\hat \Phi_x(u) = \frac{1}{\lambda^{2-\delta} u^{\delta-1}}.
\end{equation}
Anti-Laplace transforming Eq. (\ref{antido3}), we obtain 
\begin{equation}
\Phi_{x}(t) = \frac{1}{\Gamma(\delta-1) \lambda^{2-\delta} t^{2-\delta}}.
\end{equation}
We interpret $2- \delta$, which fits the condition $0 < 2 - \delta < 1$, as the IPL index of a FGN corresponding to a Hurst coefficient $H'$ different from $H$, using 
\begin{equation} \label{newdelta}
\delta' = 2 - \delta.
\end{equation}
In other words using for both $H$ and $H'$ Eq. (\ref{cakirrelation}), more precisely, $H = 1 -\delta/2$  and $H' = 1 -\delta'/2$, 
we obtain Eq. (\ref{veryimportant}).
We stress that to realize the condition of Eq. (\ref{cakirrelation}) the coupling $\Delta$ must be large enough as to annihilate the stretched exponential regime $t < 1/\lambda$. Weak values of $\Delta$, as we see in
Section \ref{numerical1}, in addition to an extended stretched exponential regime of the regression to equilibrium of $\left<x(t)\right>$ generate an extended time regime where the single trajectories return to the origin
with a IPL waiting time pdf. 

\subsection{Moving from super-diffusion} \label{movingfromsuperdiffusion}
In this case we set
\begin{equation}
\Phi_{\xi}(t) = \frac{\gamma}{\left(1 + \gamma t\right)^{\delta}}
\end{equation}
thereby yielding the asymptotic limit
\begin{equation} \label{contiunuoustime4}
\mathcal{L} \left\{\Phi_{\xi}(t)\right\} = \frac{\Gamma(1-\delta)}{u^{1-\delta}}.
\end{equation}
Following the same approach as that adopted in the preceding subsection, we obtain for the Laplace transform of the correlation function of $x$ the following expression
\begin{equation} \label{rohishasuggestion3}
\hat \Phi_x(u) = \frac{1}{u+ \Delta^2 \Gamma(1-\delta) u^{\delta-1}} .
\end{equation}
In the limiting case of $\Delta$ very large 
\begin{equation} \label{rohishasuggestion4}
\hat \Phi_x(u) = \frac{1}{\Delta^2 \Gamma(1-\delta) u^{\delta-1}} 
\end{equation}
and in time regime we obtain, using again $\alpha = 2 -\delta$, which in this case makes $\alpha > 1$, 
\begin{equation} \label{rohishasuggestion5}
\Phi_{x}(t) =  \frac{1}{\Gamma(\alpha -1)\Delta^{2} } \frac{1}{\Gamma(1-\alpha)} \frac{1}{t^{\alpha}}.
\end{equation}
Using the well known relation 
\begin{equation}
\Gamma(1-z) \Gamma(z) = \frac{\pi}{\sin \pi z},
\end{equation}
with $z$ being a generic real number, we can rewrite Eq. (\ref{rohishasuggestion5}) as
\begin{equation} \label{rohishasuggestion6}
\Phi_{x}(t) =   - \frac{(1-\alpha)}{\Delta^{2}}\frac{\sin(\pi \alpha)}{\pi}  \frac{1}{t^{\alpha}}.
\end{equation}
Notice that the continuous time representation of Eq. (\ref{rohishasuggestion3}) would yield $\hat \Phi_x(0) = 0$ in full agreement with the localization condition of Eq. (\ref{sub}). We trust that the discrete time representation adopted by the numerical treatment of  Section \ref{numerical1} with the  normalization condition 
$\Phi_{\xi}(0) = 1$ establishes an abrupt drop of this initial condition to the negative tail of Eq. (\ref{rohishasuggestion5}). Interpreting $\alpha$ as the power index $\delta'$ of $H' < 0.5$,  and using again Eq. (\ref{newdelta}) we recover
Eq. (\ref{veryimportant}).

\section{Renewal event approach to the time convoluted Fokker-Planck equation} \label{thesameinitialcondition}

In this Section we derive the ML relaxation using subordination. To realize this goal we derive a time convoluted Fokker-Plank equation with the same structure as Eq. (\ref{timeconvoluted}), moving from the operational time $n$ to the clock time $t$. We have shown that Eq. (\ref{timeconvoluted}) generates for the relaxation to equilibrium Eq. (\ref{ensemble1}), identical to GLE approach, and consequently to the ML relaxation of Eq. (\ref{ml}).  

The operational time is discrete and it  is made equivalent to a continuous time  by replacing the Fokker-Planck operator of Eq. (\ref{definition1}) with 
\begin{equation} \label{definition2}
\mathcal{L}_{FP}  \equiv  \omega \left\{\frac{d}{d x}x + \left<x^2\right>_{eq} \frac{d^{2}}{d x^{2}}\right\} ,
 \end{equation}
 where
 \begin{equation} \label{crucialcondition}
 \omega \ll 1.
 \end{equation}
 Let us adopt the CTRW perspective, \cite{montrollweiss,pramukkul}
\begin{equation} \label{ctrw2}
p(x,t) = \sum_{n = 0}^{\infty} \int_{0}^{t} dt' \psi_{n}(t') \Psi(t-t') \left[exp({\mathcal{L}_{FP}  n}) p(x,0)\right],
\end{equation}
where $\psi_n(t)$ is the probability that an event occurs at $t$ for the $n$-th time. 
The crucial condition (\ref{crucialcondition}) makes very large the number of events $n$ necessary to generate significant fluctuation-dissipation changes.  When very large values of $n$ are involved, we can interpret $n$ as a continuous \emph{dimensionless} time. In the literature  the continuous time $n$ is
usually termed as \emph{operational} time \cite{operational}. 
However, to make our model more attractive with an  anthropomorphic metaphor we assume \cite{psychological}  that 
the subjective time of the runner does not coincide with the clock time and we refer to the continuous limit of $n$ as ``psychological" time. Another interpretation is that the runner between two consecutive actions
is sleeping, an anthropomorphic metaphor corresponding to the trapping  of the diffusing molecule.  

The adoption of discrete time representation allows us to interpret the process as resulting from the occurrence of renewal events. The time distance between two consecutive renewal events is driven by the waiting time PDF $\psi(\tau)$, which is either derived from the idealized Manneville map \cite{compression} or from the waiting time PDF associated to the Mittag-Leffler function \cite{scalas}. Both pictures generate a survival probability $\Psi(t)$ with the time asymptotic property
\begin{equation}
\lim_{t \rightarrow \infty}\Psi(t) = \left(\frac{T}{t}\right)^{\mu-1}
\end{equation}
and the waiting distribution density
\begin{equation} \label{futurework}
\lim_{t \rightarrow \infty} \psi(t) = \frac{(\mu-1)T^{\mu-1}}{t^{\mu}},
\end{equation}
with
\begin{equation}
\mu < 2. 
\end{equation}
Notice that the index $\mu$ adopted to define the subordination procedure must not be confused with the complexity index $\mu_R$ defined  in Eq. (\ref{complexity}).  The IPL index $\mu_R$ is a property of the return to the origin of the single trajectories in both the subordination and the GLE case. The IPL index $\mu$ refers to the crucial events 
adopted to define the subordination process. It is important, in fact, to reiterate that we call \emph{crucial events}  the operational time renewal events occurring in the clock time.  

It is straightforward to prove, adapting to this case the  algebra illustrated in  Ref. \cite{onsager},   that Eq. (\ref{ctrw2}) is equivalent to the time convoluted form 
\begin{equation} \label{kenkrewrong}
\frac{d}{d t} p(x,t) = \int_{0}^{t} dt' \varphi_{MW}(t-t') \left(exp(\mathcal{L}_{FP})-1)\right)  p(x,t'),
\end{equation}
which becomes identical to
\begin{equation} \label{kenkrewrong2}
\frac{d}{d t} p(x,t) = \int_{0}^{t} dt' \varphi_{MW}(t-t') \mathcal{L}_{FP}  p(x,t'),
\end{equation} 
due to the condition  (\ref{crucialcondition}).

Note that $\varphi_{MW}(t)$ is the Montroll-Weiss memory kernel defined through its Laplace transform by
\begin{equation}\hat \varphi_{MW}(u) =  \frac{u \hat \psi(u)}{1 - \hat \psi(u)}.
\end{equation}

Let us now establish a connection between Eq. (\ref{kenkrewrong2}) and Eq. (\ref{ensemble1}). To do that let us evaluate the time evolution of $\left<x(t)\right>$ using Eq. (\ref{kenkrewrong2}). 
We get
\begin{multline} \label{rohisha}
\frac{d\left<x(t)\right>}{dt} =  \frac{d}{dt} \int_{-\infty}^{+\infty} dx x p(x,t) \\ =  \int_{0}^{t} dt  ' \varphi_{MW}(t-t') \int_{-\infty}^{+\infty} dx x   \mathcal{L}_{FP}  p(x,t).
\end{multline}

To get this result we made the assumption that the time derivative appearing in the second term of Eq. (\ref{rohisha}) commutes with the integral over $x$ so as to apply Eq. (\ref{kenkrewrong2}). Then we made the assumption that the time integral commutes with the integral over $x$. By applying the operator  $ \mathcal{L}_{FP}$ to $x$ using the method of integration by parts  and taking into account that the second order derivative of this operator applied to $x$ yields a vanishing value, we get
\begin{equation} \label{rohisha2}
\frac{d\left<x(t)\right>}{dt} =  - \omega \int_{0}^{t} dt  ' \varphi_{MW}(t-t') \left<x(t')\right> .
\end{equation}
This equation becomes identical to Eq. (\ref{ensemble1}) by either setting
\begin{equation} \label{rohisha3}
\omega \varphi_{MW}(t) = \Delta^2 \Phi_{\xi}(t)
\end{equation}
or, equivalently,
\begin{equation} \label{rohisha4}
 \varphi_{MW}(t) = \Omega^2 \Phi_{\xi}(t),
\end{equation}
where
\begin{equation}
\Omega^2 \equiv \frac{\Delta^2}{\omega}.
\end{equation}
We remind the readers that $\omega$ is dimensionless. 

The important conclusion of this Section is that the relaxation function
\begin{equation}
G(t) \equiv \frac{\left<x(t) \right>}{\left<x(0)\right>},
\end{equation}
generated by subordination becomes identical to FBM relaxation of Eq. (\ref{ml}) when  Eq. (\ref{rohisha4}) applies.  In fact, in this case $\omega  \varphi_{MW}(t)$ of Eq. (\ref{rohisha2}), due to 
(\ref{rohisha4}), becomes identical to the memory kernel $\varphi(t)$ of Eq. (\ref{fromphitoPhi}),  which leads to Eq. (\ref{ml}).

\section{Single trajectory behavior} \label{trajectories}

In this Section we analyze the time evolution of the single trajectories corresponding to the ensemble treatment of the earlier Sections. 

\subsection{Single  GLE trajectories } \label{numerical1}
In this Section we run the GLE of Eq. (\ref{gle}) with the memory kernel $\varphi(t)$ given by Eq. (\ref{fromphitoPhi}).

\begin{figure} [H]
 	\begin{center}
		\includegraphics[width=0.9\linewidth]{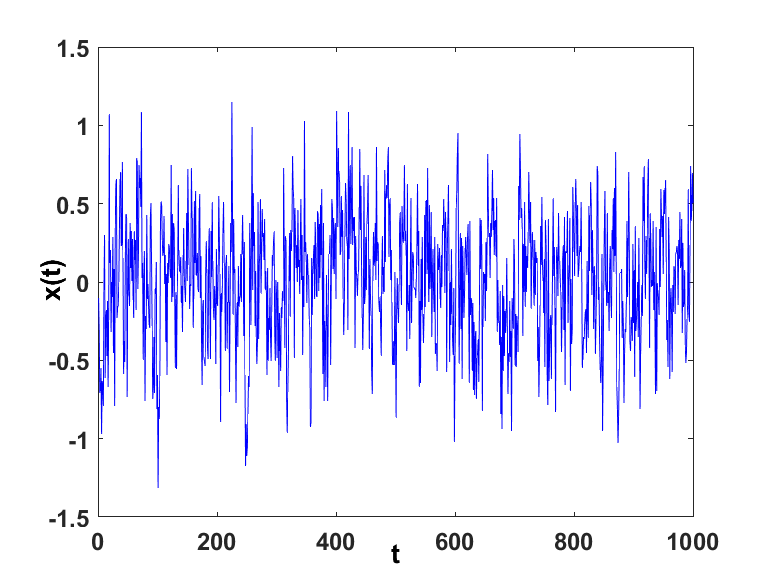} 
		\caption{Time evolution of $x(t)$ driven by Eq. (\ref{gle}) with the parameters $H=0.25$ and $\Delta^2=0.3$.}
		\label{FIG0}
	\end{center}
 \end{figure}
  
The trajectory $x(t)$ illustrated in Fig. \ref{FIG0} is an example of the fractional trajectories that we study in this Section. 
To make a quantitative analysis of  the fractal properties of their time evolution, we 
detect the time distance between two consecutive origin crossings and we evaluate the corresponding survival probability.  We apply this approach to these trajectories  
 for different values of the coupling parameter $\Delta$. 

In the limiting case $\Delta = 0$, the regression to the origin is exactly the same as that generated by FBM \cite{fabiomauro}. The theory of this paper yields the following time asymptotic expression for the 
waiting time PDF
\begin{equation} \label{greatMAUROgreat}
\psi(t) =  \frac{C_1}{t^{2-H}} +\frac{C_2}{t^{1 + 2H}}.
\end{equation}
The Fractional Gaussian noise $\xi(t)$ generating the stationary correlation function of Eq. (\ref{wanted}) was found using the FBM algorithm of Ref. \cite{elvis}.
This leads us to the analytical formula for the corresponding survival probability:
\begin{equation} \label{superposition}
\Psi(t) = \frac{1-c}{(1+t)^{{1-H}}} + \frac{c}{(1+t)^{{2H}}},
\end{equation}
where $c$ is the fitting parameter. Notice that the authors of Ref. \cite{fabiomauro} studied the asymptotic time limit and proved that, if $H < 1/3$, the complexity index $\mu_R$ is given by:
\begin{equation} \label{non-renewal}
\mu_R = 1 + 2 H
\end{equation}
and if $H > 1/3$ it is given by:
\begin{equation} \label{renewal2}
\mu _R = 2 - H.
\end{equation}

The numerical treatment of this paper in the presence of an even small value of $\Delta^2$ prevents us from exploring this asymptotic time regime, thereby making it difficult for us to see the emergence of either Eq. (\ref{non-renewal}), for $H < 1/3$, or of Eq. (\ref{renewal2} ), for $H < 1/3$. This distinction becomes evident for values of $H$ significantly larger than $1/3$. 
If  $\Delta^2$ does not vanish, but it is very weak, the generalized non-Markov friction is not yet strong enough as to cancel any sign of free regression to the origin, and consequently any sign of Eq. (\ref{non-renewal}) and Eq. ({\ref{renewal2}). In the long-time region, however, as an effect of non-Markovian friction, we observe the emergence of an exponential truncation.

To get a better understanding of this exponential truncation we increase the value of $\Delta$ so that  according to Eq. (\ref{lambdadefined}) the value of the parameter $\lambda$ of the ML survival probability is 
$\approx 1$ . We remind the readers that the stretched exponential of the ML survival probability appears in the region $t < 1/\lambda$. Due to our choice of $\Delta t = 1$, this condition implies that no sign of the initial stretched exponential is allowed to appear. Consequently, we reach the conclusion that increasing the intensity of the non-Markov friction has the effect of turning the ML survival probability into an IPL. 
In this condition the variable $x$ becomes exactly identical to a fractional Gauss noise with $H' = 1-H$.
According to a theorem established by the authors of Ref. \cite{rohishatheorem} the waiting time distribution of the time distance between two consecutive origin crossings is given by 
\begin{equation} \label{rohishaexponential}
\psi(t) = r exp(-rt), 
\end{equation}
where
\begin{equation} \label{rohisharate}
r = 1 - \frac{2}{\pi} arcsin 2^{H-1}. 
\end{equation}

\begin{figure}
 	\begin{center}
		\includegraphics[width=0.9\linewidth]{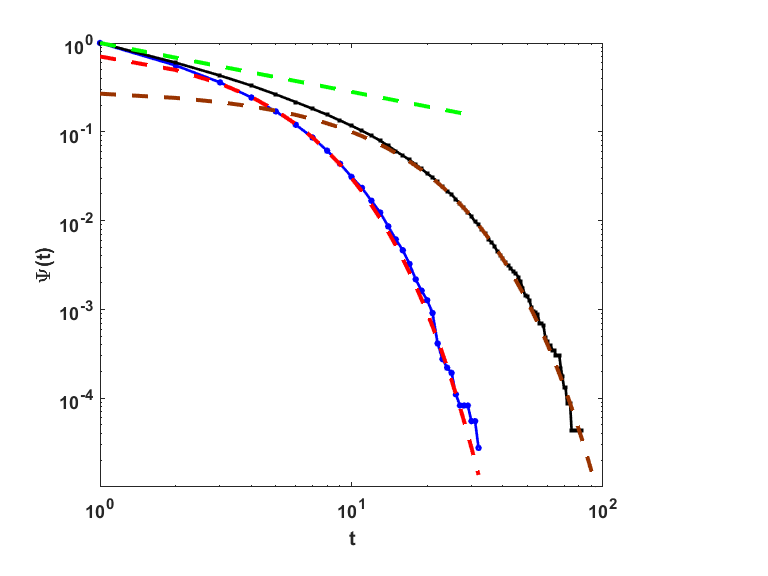} 
	
\caption{Cumulative probability for the recrossing of the origin of $x(t)$ of Eq.(\ref{gle}) for different values of $\Delta$ and $H = 0.25$.  The black curve refers to the case $\Delta^2 = 0.08$ and the blue curve refers to $\Delta^2 = 0.3$. The top green dashed line illustrates the slope predicted by Eq. (\ref{non-renewal}) with $\mu_R-1=0.5$. The brown dashed line is the fitting given by $e^{-0.11 t}$ which is the truncation expected in the long time limit. The red dashed line is the fitting $e^{-0.35 t}$, which shows the exponential waiting time PDF of Eq. (\ref{rohishaexponential}) with $H$ of Eq. (\ref{rohisharate}) replaced by $H'= 1-H$. }
		\label{FIG1}
	\end{center}
 \end{figure}

These properties are illustrated in 
 Fig. (\ref{FIG1}). We see that for $\Delta \approx 0.28$ the survival probability is an IPL with an exponential truncation.
These are properties of the condition $\Delta^2 = 0$ that remain present also with a non-vanishing value of $\Delta^2$ if this is sufficiently small.  We should observe the slowest scaling  of Eq. (\ref{greatMAUROgreat}), that in the case $H < 1/3$ is  $\mu_R$ of Eq. (\ref{non-renewal}). 
However, in the long-time region it cannot show up as an effect of the friction-induced exponential truncation, and it is confined to the  short-time region,  as shown by Fig. \ref{FIG1}.  The long-time limit is characterized by an exponential truncation that has a rate $r$ smaller than the rate generated by a large friction making the exponential relaxation predominant, $r = 0.11$ versus $r= 0.35$.  
When $\Delta \approx 0.55$ the exponential truncation becomes predominant  and extends from the short to the long-time regime. In this case  the non-Markov friction is large enough as to turn the ML relaxation into a mere IPL, so that for the single trajectories the recrossing of the origin fits very well the prediction 
of Eqs. (\ref{rohishaexponential}) and (\ref{rohisharate}). 

In order to establish a contrast with the discussion of Section \ref{numerical2} we study the condition of  Eq. (\ref{renewal2}) (see Fig. (\ref{FIG3})). In this case  the slowest contribution to Eq. (\ref{greatMAUROgreat}) is  the IPL of Eq. (\ref{renewal2}), which is visible. This is so because it yields the slope $1.6$, slightly larger than the corresponding value  $1.5$ of Fig. (\ref{FIG1}). In fact, the faster IPL decay is expected to show up at times short enough as to be still unaffected
by the friction-induced exponential truncation.  
\begin{figure}
		\includegraphics[width=1\linewidth]{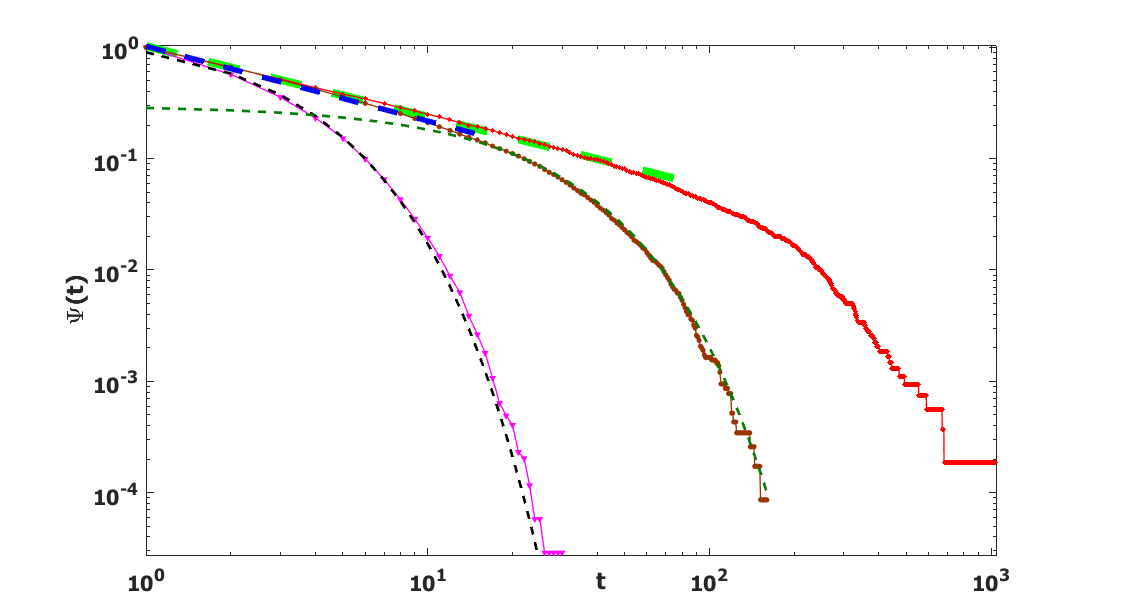} 
\caption{Cumulative probability for the recrossing of the origin of $x(t)$ of Eq.(\ref{gle}) for different values of $\Delta$ and $H = 0.4$.  The three curves, moving from the top down, refer to $\Delta^2 = 0.009, 0.05, 0.5$. The dashed line of the top curve illustrates the slope corresponding to  Eq. (\ref{renewal2}), namely $\mu_R - 1 = 1 -H$. The dashed line of the middle curve shows an exponential truncation in the long time limit. The dashed line of the bottom curve corresponds to the exponential waiting time PDF of Eq. (\ref{rohishaexponential}) with $H$ of Eq. (\ref{rohisharate}) replaced by $H'= 1-H$. }
		\label{FIG3}
 \end{figure}

 \begin{figure} 
 	\begin{center}
		\includegraphics[width=1\linewidth]{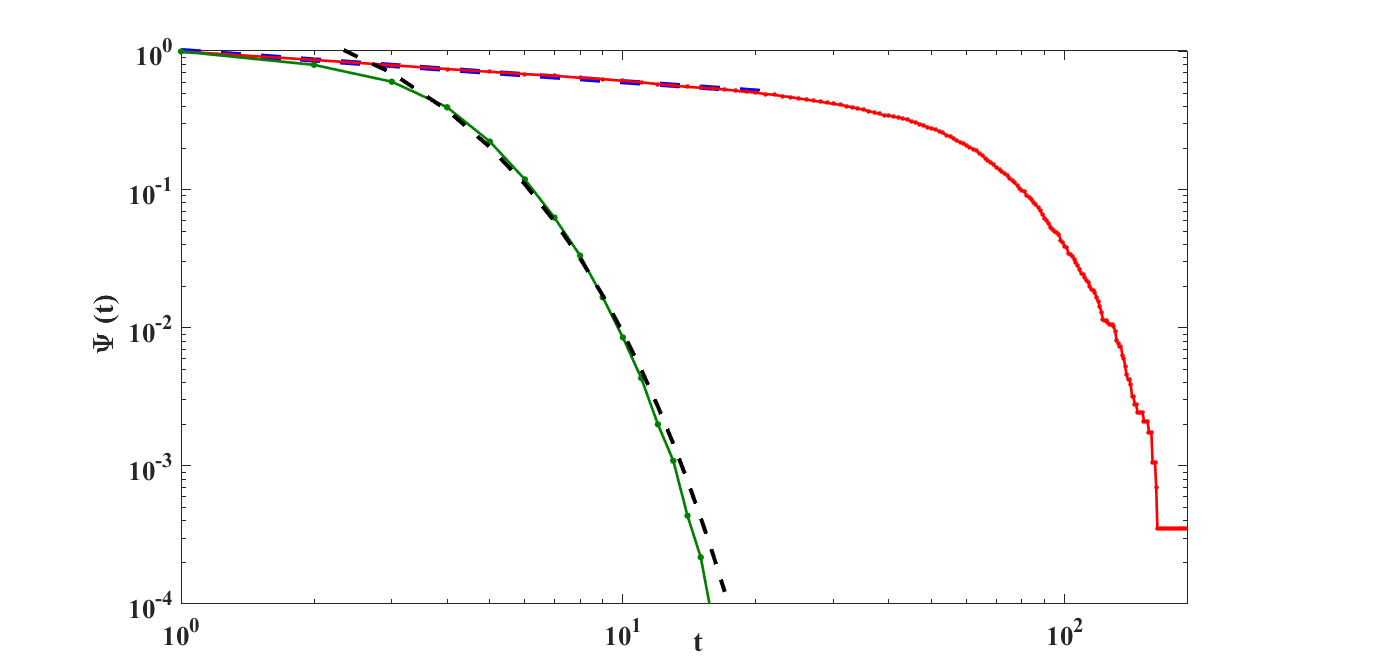} 
		\caption{Cumulative probability for the recrossing of the origin of $x(t)$ of Eq.(\ref{gle}) for different values of $\Delta$ and $H = 0.8$.  The red curve refers to $\Delta^2 = 0.009$ and the black curve refers to $\Delta^2 = 0.61$.  The blue dashed line is the slope corresponding to  Eq. (\ref{renewal2}), namely $\mu_R - 1 = 1 -H \approx 0.22$. The black dashed line is the exponential waiting time PDF of Eq. (\ref{rohishaexponential}) with $H$ of Eq. (\ref{rohisharate}) replaced by $H'= 1-H$.}
		\label{FIG4}
	\end{center}
 \end{figure}
 
 The result of Fig. (\ref{FIG4}) is impressive. In fact, it refers to the condition studied in Section \ref{movingfromsuperdiffusion}, which corresponds to generate a ML function with $\alpha > 1$. 
 According to the theoretical arguments of this earlier Section in the limiting case of strong friction, the variable $x$ is expected to become equivalent to the fractional Gaussian noise generating subdiffusion.
 The numerical results of this figure fully confirms that prediction.

\subsection{Single trajectories according to the subordination perspective} \label{numerical2}
The regression of $<x(t)>$ to the vanishing mean value is fairly described by the theory of Section \ref{thesameinitialcondition}. Here we focus our attention on the reiterated regression to the origin of the single trajectories. We prove that in this case the long-time limit the cumulative probability of the trajectory regression to the origin is always a power law, in  a deep contrast with the numerical and theoretical results of Section \ref{numerical1}, where the cumulative probability has always an exponential long-time limit behavior.  In the case of $\omega$ very large but not exceeding 1 we have 
\begin{equation} \label{large}
\mu_R = \mu
\end{equation}
and in the case of virtually vanishing value of $\omega$ we
have 
\begin{equation} \label{small}
\mu_R = (1+\mu)/2.
\end{equation}
We prove  Eq. (\ref{large}) and Eq. (\ref{small}) theoretically and 
we double check these limiting conditions numerically. Using a numerical treatment we study also an intermediate case. 

Let us prove Eq. (\ref{large}) first. In the operational time $n$
the single trajectories are described by the ordinary Langevin equation
\begin{equation} \label{le2}
\frac{d}{dn} x = - \omega x(n) + \xi(n),
\end{equation}
where $\xi(n)$ is either $1$ or $-1$, according to a fair coin tossing. The operational time $t$ is discrete, but we study the long-time limit $n \gg 1$, which makes us interpret it as a continuous time with the elementary time step $\Delta n = 1$. In the case where $\omega$ is of the order of $1$ 
the time distance between two consecutive origin recrossing in the operational time scale is of the order of $1$ and consequently indistinguishable from the coin tossing process, which yields for the survival probability $\Psi(n)$ the prescription
\begin{equation}
\Psi(n) = \left(\frac{1}{2}\right)^n = exp(- n \cdot ln 2). 
\end{equation}
We are immediately led to conclude that in the clock time this
survival probability becomes
\begin{equation}
\Psi(t) \propto \frac{1}{t^{(\mu-1)}},
\end{equation}
with the waiting time PDF $\psi(t)$ given by
\begin{equation}
\psi(t) \propto \frac{1}{t^{\mu}}.
\end{equation}
This proves Eq. (\ref{large}).   In the next subSection in addition to proving Eq.(\ref{small}) we afford an alternate proof of Eq. (\ref{large}). 

To prepare the ground for the demonstration of Eq. (\ref{small}), done in the next sub-Section, let us refer Fokker-Planck equation of Eq. (\ref{fp}) to the operational time $n$  using the following equation
\begin{equation} \label{rohishaimprovingnotation}
\frac{\partial}{\partial n} p(x,n) = \left( \omega \frac{\partial}{\partial x} x + D \frac{\partial^2}{\partial x^2} \right) p(x,n).
\end{equation}
In this case we  have $D = \omega \left<\ x^2 \right >_{eq}$.  When friction is very small, $\omega \ll D$, we have $1 \ll \left<\ x^2 \right >_{eq}$ and the friction term can be neglected compared to the diffusion term.
This would lead to an equation with only the diffusion term, like Eq. (\ref{rohishanotation}). Note that the time dependent diffusion coefficient $Q(t)$ of Eq. (\ref{rohishanotation}) becomes independent of time and identical to $D$ when $H = 0.5$.

\subsection{Single trajectories according to the subordination perspective in the extreme case of no friction}
Let us now move to discuss the case of $\omega$  so small as to disregard the friction term as mentioned earlier, namely, for simplicity's sake let us set $\omega = 0$ in Eq. (\ref{rohishaimprovingnotation}). 
In this case the diffusion process becomes identical to the popular Continuous Time Random Walk(CTRW) \cite{montrollweiss,pramukkul}. The CTRW pdf p(x, t) can be written as \cite{pramukkul}
\begin{equation}\label{fde1}
 \frac{\partial^{\alpha}}{\partial t^{\alpha}}p(x,t)=D  \frac{\partial^{2}}{\partial x^{2}}p(x,t),
\end{equation}
where
\begin{equation} \label{definition}
\alpha \equiv \mu -1.
\end{equation}
According to Ref. \cite{pramukkul}, the fractional derivative on the left hand side of Eq. (\ref{fde1}) is a Caputo's fractional derivative. 

To deal with Eq. (\ref{fde1}) is convenient to use the Fourier-Laplace transform method. The Fourier-Laplace transform of the function $f(x,t)$ is defined by the
notation
\begin{equation}
\hat{\hat{f}} (k,u) \equiv  \int_{-\infty}^{\infty}dx e^{ikx}\int_{0}^{\infty}dt e^{-ut} f(x,t).
\end{equation}
To simplify this heavy notation, from here on we do not use the double hat, and we do not use the single hat either, and we adopt the convention that
$f(k,u)$ denotes the Fourier-Laplace transform of $f(x,t)$ and $f(x,u)$ its inverse Laplace transform.

Performing the Laplace-Fourier transform we have
\begin{eqnarray}\label{fde2}
    {p}(k,u) = \frac{ u^{\alpha-1}}{ u^{\alpha}+D k^2},
\end{eqnarray}
where we assumed that $p(x,0)=\delta(x)$. Going back to the $x$ space we have

\begin{equation}\label{fde3}
 {p}(x,u) = \frac{u^{\frac{\alpha }{2}-1} e^{-\frac{\left| x\right|  u^{\alpha /2}}{\sqrt{D}}}}{2 \sqrt{D}}.
\end{equation}

The adoption of the prescription of Ref. \cite{redner} to establish the first-passage time to get the origin moving from the origin cannot be adopted, because it rests on the assumption  that the single trajectory leaves the origin immediately with no extended resting time on it.  This assumption violates the condition established by the subordination approach that may make the particle rest for a long time on the origin. To evaluate the first time for the regression to the origin it is convenient to consider a  strip of small size $\epsilon$ around the origin and to evaluate the time necessary for the trajectory to re-enter this stripe after leaving the origin, without forcing the event of leaving the origin to  occur with no delay.  In other words, we evaluate the first passage time from $0$ to $x >0$, where $x = \epsilon \ll 1$.  To make the readers aware of the adoption  of this procedure to establish the origin to origin regression, we denote the waiting time  PDF with the symbol 
$\psi_x(t)$ rather than $\psi(t)$. According to  Ref. \cite{redner} we have that
the first-passage time distribution ${p}(x,u)$ is related to the density $p( x,t )$ via  the relation

\begin{eqnarray}\label{fde4}
{p}(x,u) = {p}(0,u)  \hat{\psi}_{x}(u).
\end{eqnarray}
We may rewrite Eq. (\ref{fde4}) as

 \begin{equation}\label{fde5}
 \hat{\psi}_{x}(u)=e^{-\frac{\left| x\right|  u^{\alpha /2}}{\sqrt{D}}}
\end{equation}
that is the one-sided L\'evy distribution. Using the Bromwich contour we may write for $\psi_{x}(t)$ the expression:

 \begin{align}
\psi_{x}(t)&=\frac{1}{\pi }
\int_0^{\infty } \exp\left[-\frac{\left| x\right|}{\sqrt{D}}  u^{\frac{\alpha }{2}} \cos\frac{\pi \alpha }{2}-tu\right]\cdot \nonumber\\
& \sin \left[\frac{\left| x\right|}{\sqrt{D}} u^{\frac{\alpha }{2}} \sin \frac{\pi \alpha}{2} \right]du.
\label{fde6}
\end{align}
The asymptotic behavior can be deduced from Eq. (\ref{fde5})

 \begin{equation}\label{fde7}
 \hat{\psi}_{x}(u)\approx 1-\frac{\left| x\right|  u^{\alpha /2}}{\sqrt{D}}
 \end{equation}
corresponding to

  \begin{equation}\label{fde8}
 \psi_{x}(t)\approx -\frac{\left| x\right| }{\Gamma\left[-\frac{\alpha }{2}\right]\sqrt{D} t^{\frac{\alpha }{2}+1}}. 
 \end{equation}
 This is equivalent to setting $\mu_R =  \alpha/2 + 1$, which, taking into account $\alpha \equiv \mu-1$ yields $\mu_R = (1+\mu)/2$, identical to Eq. (\ref{small}). This is the proof that the lack of a friction yields Eq. (\ref{small}).
 
\subsection{Single trajectories according to the subordination perspective with friction}
In line with Eq. (\ref{introduction2}) we study the equation
\begin{equation}\label{ffde1}
 \frac{\partial^{\alpha}}{\partial t^{\alpha}}p(x,t)=D  \frac{\partial^{2}}{\partial x^{2}}p(x,t)+\omega\frac{\partial}{\partial x}\left[xp(x,t)\right].
\end{equation}
Taking the Fourier-Laplace transform we have

\begin{equation}\label{ffde2}
 u^{\alpha } p(k,u)-u^{\alpha -1}=-D k^2 p(k,u)-\omega  k \frac{\partial p(k,u)}{\partial k}.
 \end{equation}
The solution of the above equation is

\begin{align}
 p(k,u)&= c \exp\left[-\frac{D k^2}{2 \omega }\right] |k|^{-\frac{u^{\alpha }}{\omega }} -\frac{u^{\alpha -1} }{2 \omega } 
 \exp\left[-\frac{D k^2}{2 \omega }\right]\cdot \nonumber\\ & Ei_{1-\frac{u^{\alpha }}{2 \omega }}\left(-\frac{D k^2}{2 \omega }\right), 
\label{ffde3}
 \end{align}
 where $Ei_a(z)$ is the exponential integral function.  The constant $c$ has to be chosen in such a way that $p(0,u)=1/u$. What is left is
 
\begin{equation}\label{ffde4}
 p(k,u)=\sum_{n=0 }^\infty \frac{ 2^{-n} D^n \omega ^{-n} k^{2 n}}{n!}
\frac{u^{\alpha -1}}{u^{\alpha }+2 \omega  n}\exp\left[-\frac{D k^2}{2 \omega }\right] .
 \end{equation}
  Inverting the Fourier transform we obtain
 
\begin{equation}\label{ffde5}
 p(x,u)=\sqrt{\frac{ \omega}{2 D \pi^2} }
 \sum _{n=0}^{\infty } \frac{ F_{\Gamma}(n, x)}{ n! } \frac{u^{\alpha -1}}{u^{\alpha }+2 \omega  n},
 \end{equation}
 where  
 \begin{equation}
 F_{\Gamma}(n, x) \equiv  \Gamma \left(n+\frac{1}{2}\right) F_1\left(n+\frac{1}{2};\frac{1}{2};-\frac{x^2 \omega }{2 D}\right)
 \end{equation}
 and 
 $ F_1\left(a;b;z\right)$ is the confluent hypergeometric function.
 
  Finally inverting the Laplace transform we end up into

\begin{equation}\label{ffde6}
 p(x,t)=\sqrt{\frac{ \omega}{2 D \pi^2} }
 \sum _{n=0}^{\infty } \frac{ F_{\Gamma}(n, x)}{ n! } E_{\alpha}\left(-2\omega n t^{\alpha}\right),
 \end{equation}
where $ E_{\alpha}\left(z\right)$ is the Mittag-Leffler function. Note that being the first argument of the hypergeometric an half-integer, then  $ F_1\left(n+\frac{1}{2};\frac{1}{2};-\frac{x^2 \omega }{2 D}\right) $  can be written as

 \begin{equation}\label{ffde7}
 F_1\left(n+\frac{1}{2};\frac{1}{2};-\frac{x^2 \omega }{2 D}\right)=   
 \exp\left[-\frac{\omega  x^2}{2 D}\right] 
  \frac{(-1)^n   H_{2 n}\left(\sqrt{\frac{ \omega}{2  D} }x\right)}{2^n (2 n-1)!!},
  \end{equation}
where $H_n\left(z\right)$ are the ``physicists" Hermite polynomials of $n$ degree. In particular for $n=0$ we have 
  
\begin{equation}\label{ffde8}
 F_1\left(\frac{1}{2};\frac{1}{2};-\frac{x^2 \omega }{2 D}\right)=   
 \exp\left[-\frac{\omega  x^2}{2 D}\right]  
 \end{equation}
so that the first term of series (\ref{ffde6}) is the equilibrium distribution, i.e.
  
\begin{equation}\label{ffde9}
  p_{eq}(x)=   \sqrt{\frac{ \omega}{2\pi D} }\exp\left[-\frac{\omega  x^2}{2 D}\right]  .
 \end{equation} 
 The Laplace transform of the first-passage time distribution $ \psi_{x}(t)$ is given by  

\begin{eqnarray}\nonumber
 \hat{\psi}_{x}(u)=\frac{{p}(x,u) }{{p}(0,u)}&=&\sqrt{\frac{ \omega}{2 D \pi^2} }
 \sum _{n=0}^{\infty } \frac{F_{\Gamma}(n, x)}{ n! } \frac{u^{\alpha -1}}{u^{\alpha }+2 \omega  n}  \\\label{fde10}
&&\times\sqrt{\frac{2 D}{ \omega} }
  \frac{u \Gamma \left(\frac{u^{\alpha }}{2 \omega }+\frac{1}{2}\right)}{  \Gamma \left(\frac{u^{\alpha }}{2 \omega }+1\right)},
\end{eqnarray}
which  for $u\to 0$ yields
 \begin{equation}
  \hat{\psi}_{x}(u)
 \approx \exp\left[-\frac{\omega  x^2}{2 D}\right] \left[1+f(x) u^{\alpha }\right].
 \end{equation}
 
In the time representation we have
 
 \begin{eqnarray}\label{fde11}
 \psi_{x}(t)\approx \exp\left[-\frac{\omega  x^2}{2 D}\right] \frac{f(x) }{\Gamma (-\alpha )t^{\alpha +1}}.
  \end{eqnarray}
  Taking into account the definition of Eq. (\ref{definition}), we obtain $\mu_R = \mu$, this being the second demonstration of Eq. (\ref{large}).

 \subsection{Numerical results for single trajectories according to subordination perspective with and without friction.} \label{numerical3}

\begin{figure}
 	\begin{center}
		\includegraphics[width=0.9\linewidth]{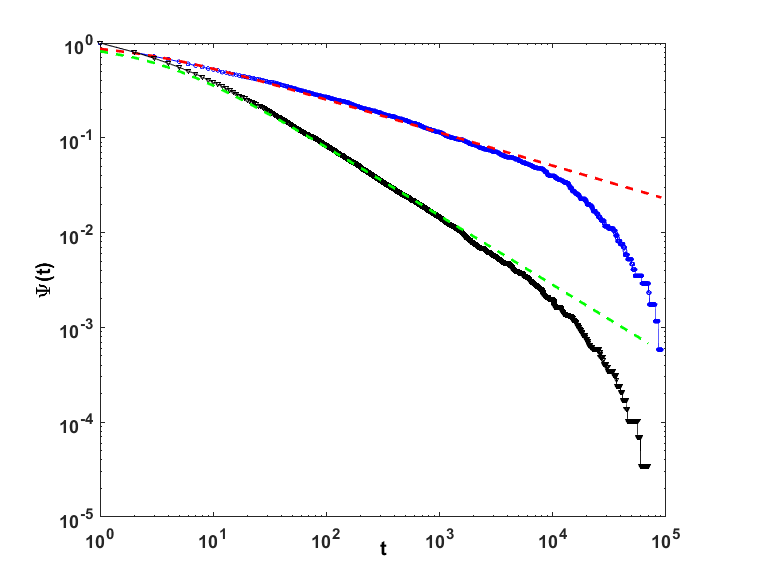} 
	
		\caption{Cumulative probability for the recrossing of the origin for different values of the fiction $\omega$. (From the top) The blue dotted line represents the numerical data for $\omega$=0.001 and the red dashed line is the fitting with scaling $\mu_R-1 \approx 0.35$; the black triangles represent numerical data for $\omega$=1 and the green dashed line is the fitting with scaling $\mu_R-1 \approx 0.73$. }
		\label{SHORTandLARGE}
	\end{center}
 \end{figure}

  The theoretical predictions of Eq. (\ref{fde8}) and Eq. (\ref{fde11}) are supported by the numerical results illustrated in Fig. (\ref{SHORTandLARGE}).  It is important to notice, however, that the exponential decay of the cumulative 
  probability $\Psi(t)$ at large times in this case is  numerical, namely, due to fact that the analysis is done using a finite length time series.

  \subsection{Important results}
  The important result of this Section is that the waiting time PDF $\psi(t)$ for the time distance of two consecutive origin crossings in the GLE case has an exponential asymptotic behavior. Making the  GLE friction intense enough has the effect of turning $\psi(t)$ into a perfect exponential. Thus the GLE case may generate an IPL behavior at short time, but it is a perfect exponential in the long-time limit, independent of the length of the observed time series.  
The subordination case  generates an exponential decay of waiting time PDF for the recursion to the origin in the long time, which is the consequence of the finite size of the observed time series. It should not be confused with the exponential truncation of the GLE theory, which is determined by the physical properties of this theory.  In the subordination  case $\psi(t)$ keeps an IPL structure with power law index $(\mu +1)/2$ for small friction and the larger power index $\mu$ for  strong friction.

 \section{CONCLUDING REMARKS} \label{last}
 
 The main result of this paper is that two different physical approaches to the ML relaxation, the former with infinite memory and no events, and the latter determined by the occurrence of crucial events, correspond to individual trajectories with a surprisingly different behavior.  The regression to the origin in the former case is described by exponential waiting time PDF and in the latter by IPL waiting time PDF.   Notice that the main focus of this paper is on the ML relaxation with $\alpha < 1$, which can derived from two totally different physical conditions, one with infinite memory and no events and the other driven by crucial events. The condition $\alpha > 1$, with the negative tail of $\Phi_{x}(t)$ is incompatible with the interpretation of this function as a survival probability. However, the subordination origin of the ML function 
 with  $\alpha > 1$ studied in the earlier work of Ref.  \cite{ascolani} does not rule out the possibility that also the correlation function generating sub-diffusion may be determined by crucial events. This is an incentive to adopt the statistical analysis of single trajectories as well as the observation of the regression to equilibrium of an out of equilibrium initial condition through the ordinary ensemble perspective. 
 
 It is important to notice that the exponential $\psi(t)$ of Eq. (\ref{rohishaexponential}) seems to conflict with the remarks in Section \ref{intro} stressing that an exponential function is considered to be incompatible with a Hamiltonian picture.
The FBM is the diffusion process generated by a  FGN and the FGN is derived \cite{cakirarkadi} from a non-Ohmic thermal bath, which has a Hamiltonian nature \cite{weiss}. However, all the arguments about the incompatibility between 
exponential relaxation and Hamiltonian treatments imply that the relaxation process is a Poisson process. As a consequence, the time distance between two consecutive crossings, $\tau_i$,  should be uncorrelated to the earlier and the later time distances. This is not true in the case of FBM derived from GLE. To stress this important fact we should evaluate the correlation function

\begin{equation} \label{rohishacorrelationfunction}
C(t) = \frac{\sum\limits_{|i-j| = t}\overline{\left(\tau_i - \overline {\tau}\right)\left(\tau_j - \overline {\tau}\right)}}{\sum\limits_i\overline{\left(\tau_i - \overline {\tau}\right)^2}}.
\end{equation}
 Proving numerically the existence of non vanishing correlation directly on the FBM derived from GLE  is hard, due to the statistical inaccuracy generated by the integration process implied by Eq. (\ref{gle}). For this reason, resting on the  theoretical arguments behind Eq. (\ref{veryimportant}), we evaluate the correlation function $C(i,j)$ of (\ref{rohishacorrelationfunction}) for two cases, one referring to  $H < 0.5$, Fig. (\ref{beautifulROHISHA1}),  and one referring to $H> 0.5$, Fig. (\ref{beautifulROHISHA2}). The departure from the renewal condition is evident. The numerical results 
of  Fig. (\ref{beautifulROHISHA1}) and Fig. (\ref{beautifulROHISHA2}), referring  to $\xi$ derived from the Mandelbrot algorithm of  Ref. \cite{algorithm}, see also Ref. \cite{elvis},  agree with the corresponding results of Ref. \cite{cakirarkadi} based on the Hamiltonian formalism of Weiss \cite{weiss}.  This supports the deep connection between FBM and Hamiltonian dynamics. 

\begin{figure}[H]
 	\begin{center}
		\includegraphics[width=0.9\linewidth]{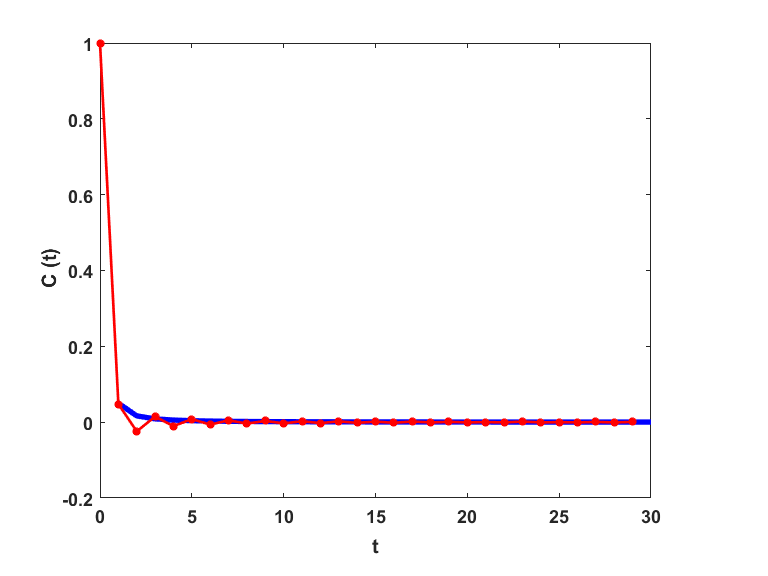} 
		\caption{Correlation function given by Eq.(\ref{rohishacorrelationfunction}) of the pernamence times of FGN with $H$=0.2. The upper blue envelope goes down as $ t^{-1.59}$.  }
		\label{beautifulROHISHA1}
	\end{center}
 \end{figure}

\begin{figure} 
\begin{center}
\includegraphics[width=0.9\linewidth]{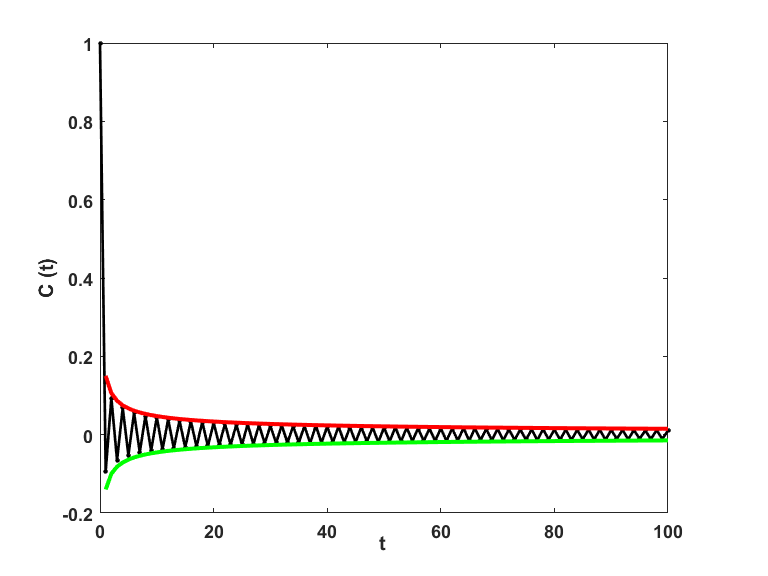} 
		\caption{Correlation function given by Eq.(\ref{rohishacorrelationfunction}) of the pernamence times of FGN with $H$=0.75. The upper red envelope goes down as $t^{-0.5}$ and the lower green envelope goes as -$t^{-0.5}$.  }
		\label{beautifulROHISHA2}
	\end{center}
 \end{figure}

  Although we have limited our attention to merely theoretical issues, the results are expected to be of interest for anomalous diffusion in crowded environments \cite{crowded,kneller} and the subject of random growth of surfaces \cite{failla}. As far as the subject of crowded environment is concerned, the adoption of fractional calculus, which is often based on the ML relaxation, must be properly connected 
 to the proper physical model involved \cite{crowded}.  The experimental observation of ergodicity breaking \cite{metzlercrowded} suggests the need of adopting the subordination picture discussed in this paper. 
 The results of computer simulation of diffusion in simple liquids suggests the adoption of the GLE picture \cite{kneller}. Both pictures lead to the same ML relaxation, thereby making it difficult to establish which is the correct model. We hope that the results of this paper, showing that the single trajectory time evolution generates   strikingly different recursions to the origin,  may help the investigators to establish the correct model to adopt.

  According to Ref. \cite{failla} subordination is expected to be a fruitful perspective to study the random growth of surfaces. However, in this field of research frequent use is made of the FBM perspective \cite{barabasi}. We hope that in this case, too, the different behavior of the single trajectories may help to establish which is the correct model to adopt with the warning, though, that in that case
  the subordination approach is realized with tempering the waiting time PDF of Eq. (\ref{futurework}) \cite{failla}.    This is an important research subject that may benefit from the results of the present paper.

\emph{Acknowledgment}| The authors thank Welch for financial support through Grant No. B-1577.


\end{document}